\documentclass[10pt,preprint,superscriptaddress,footinbib,showkeys,showpacs,tightenlines,twocolumn]{revtex4}
\usepackage{graphicx}
\usepackage{dcolumn}
\newcommand{\be}{\begin{equation}}
\newcommand{\ee}{\end{equation}}
\newcommand{\bee}{\begin{eqnarray}}
\newcommand{\eee}{\end{eqnarray}}
\newcommand{\eq}{\end{quote}}
\newcommand{\nn}{\nonumber}
\newcommand{\Slash}[1]{\ooalign{\hfil/\hfil\crcr$#1$}}

\def\lsim{\displaystyle\mathop{<}_{\sim}}
\begin{document}      
\preprint{PNU-NTG-09/2005}
\title{Suppression of $\Theta^+(J^P=3/2^{\pm})$ photoproduction from
the proton} 
\author{Seung-Il Nam}
\email{sinam@rcnp.osaka-u.ac.jp}
\affiliation{Research Center for Nuclear Physics (RCNP), Ibaraki, Osaka
567-0047, Japan}
\affiliation{Department of
Physics and Nuclear physics \& Radiation Technology Institute (NuRI),
Pusan National University, Busan 609-735, Korea} 
\author{Atsushi Hosaka}
\email{hosaka@rcnp.osaka-u.ac.jp}
\affiliation{Research Center for Nuclear Physics (RCNP), Ibaraki, Osaka
567-0047, Japan}
\author{Hyun-Chul Kim}
\email{hchkim@pusan.ac.kr}
\affiliation{Department of
Physics and Nuclear physics \& Radiation Technology Institute (NuRI),
Pusan National University, Busan 609-735, Korea} 

\date{\today}

\begin{abstract}
We investigate the photoproduction of $\Theta^+$ from the proton and
neutron, $\gamma N \to \bar K \Theta^+$.  Assuming that spin and
parity of $\Theta^+$ are  
$J^P = 3/2^{\pm}$, it is shown that the production from the proton is 
strongly suppressed as compared with that from the neutron.  
This could provide a possible explanation for the null results
of the recent CLAS experiment in finding $\Theta^+$ via the reaction
$\gamma p \to \bar K^0 \Theta^+$.   
\end{abstract}


\maketitle


Since the seminal work by Diakonov {\it et al.} 
predicted the mass and width of the
pentaquark baryon $\Theta^+$~\cite{Diakonov:1997mm}, 
a great amount of research activities has
been performed to clarify its existence and properties.  
Although many experiments reported evidences of $\Theta^+$
after the first observation by the LEPS collaboration~\cite{Nakano:2003qx},  
the situation is not yet conclusive primarily due to 
the relatively low statistics of the low energy experiments.  
Furthermore, in many of high energy experiments, the existence of
$\Theta^+$ has been doubted (see, for example, a recent
review~\cite{Hicks:2005gp} for the compilation of the experimental
results). 

Very recently, the CLAS experiment has reported no
significant evidence of $\Theta^+$ in the reaction 
$\gamma p \to \bar K^0 \Theta^+$~\cite{DeVita:2005CLAS}.  
This result should be taken seriously, 
because they have achieved significantly higher 
statistics.  
Yet their result does not lead to the absence 
of $\Theta^+$ immediately, 
because the previous positive evidences were seen mostly in 
the reactions from the neutron.  
Due to the violation of isospin symmetry in the 
electromagnetic interaction, there could be asymmetry 
in the reactions from the proton and neutron.  
A well-known example is the Kroll-Ruderman term 
in the pion photoproduction, which survives only in the charge 
exchange channels such as $\gamma p \to \pi^+ n$.  

In the present work, we would like to provide a mechanism for
the strong suppression of the reaction $\gamma p\to
\bar{K}^0\Theta^+$, as compared to $\gamma n\to {K}^-\Theta^+$.   
A similar result has been obtained in the recent work for  
$\Lambda(1520,J^P=3/2^-)(\equiv\Lambda^*)$
photoproduction~\cite{Nam:2005uq}, where we have shown the 
strong suppression of the charge non-exchange channel; 
$\sigma_{\gamma n\to K^0\Lambda^*} 
\ll \sigma_{\gamma p\to K^+\Lambda^*}$.  
The large difference between the two reactions was caused 
by the dominant contribution from the contact 
(Kroll-Ruderman like) term. 

\begin{figure}[tbh]
\resizebox{8.5cm}{4.5cm}{\includegraphics{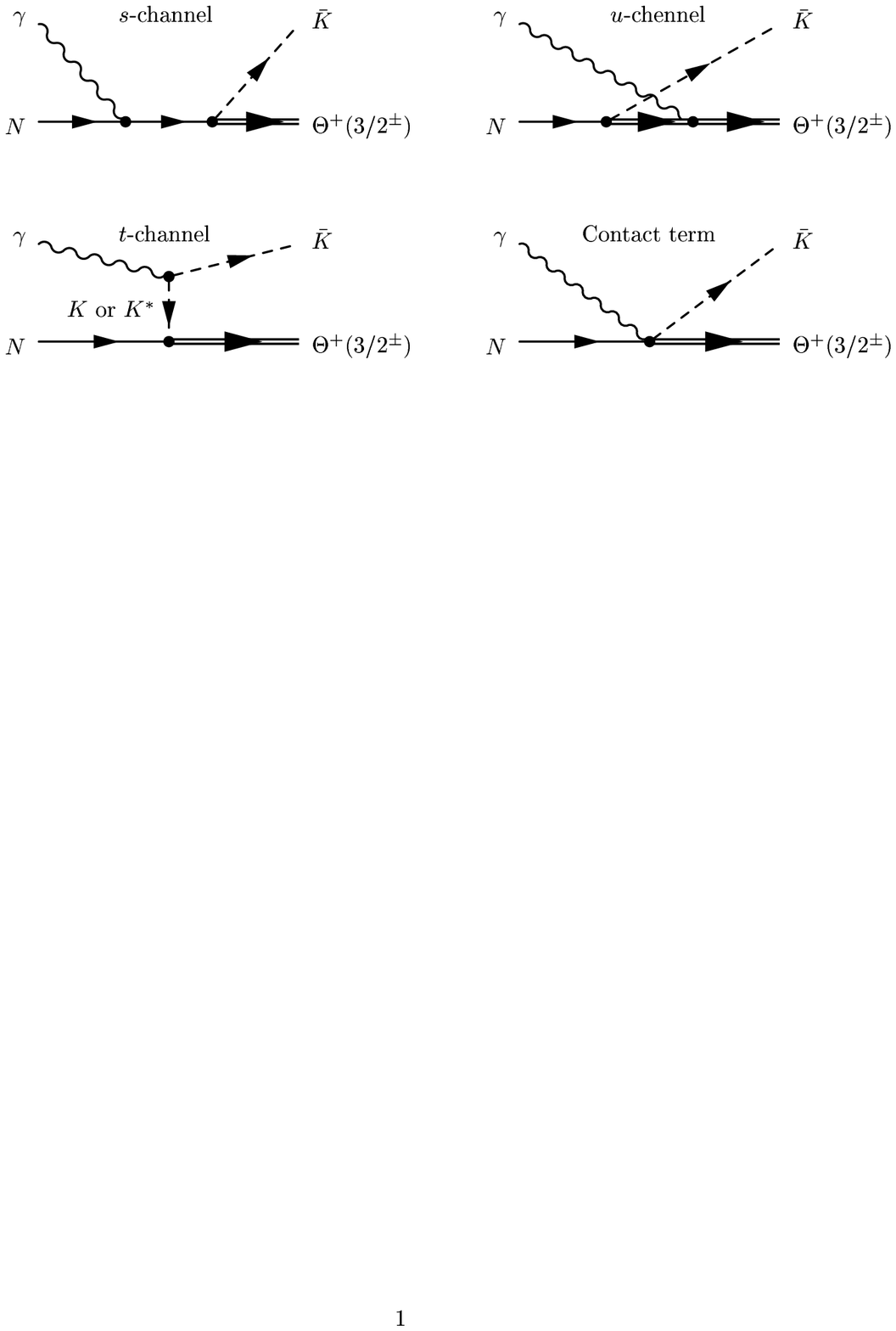}}
\caption{Born diagrams calculated in the effective Lagrangian method.}    
\label{fig00}
\end{figure}

Drawing a definite conclusion from theoretical studies 
of reactions in this energy region is rather difficult.  
However, we have accumulated empirical knowledge 
from kaon and hyperon productions, where effective 
Lagrangian method works reasonably well by choosing 
model parameters appropriately.  
In this method, the Born terms as shown in Fig.~\ref{fig00} 
are calculated.  In addition, we may add some theoretical constraints
from the low-energy side.  
Here we attempt to clarify to what extent our conclusions 
are reliable.   
Within a reasonable model setup, 
we show a remarkable role of the contact term for the 
neutron targets for $\Theta^+(J^P = 3/2^{\pm})$, which is, however, 
absent for the proton target.  
An advantage of the spin $3/2$ states especially 
with the negative parity lies in the fact that this state is
compatible with the very narrow width: in the quark model picture, the
simple configuration $(0s)^5$ forbids the decay into the $KN$ state in
the $d$-wave~\cite{Hosaka:2004bn}.    
 
Let us start with a brief description of the effective Lagrangian 
method.  
The spin 3/2 particle is treated in  
the Rarita-Schwinger formalism. 
Then the basic symmetries such as Lorentz invariance and gauge
symmetry enable one to write the interaction Lagrangians as follows:  
\bee
\mathcal{L}_{\gamma NN}
&=&-e\bar{N}\left[\Slash{A}+\frac{\kappa_{p}}{2M_{p}}
\sigma_{\mu\nu}F^{\mu\nu}\right]N\,+{\rm
h.c.},\nn\\
\mathcal{L}_{\gamma KK}&=&ie\left[(\partial^{\mu}K^{\dagger})
K-(\partial^{\mu}K\right)K^{\dagger}]A_{\mu},\nn\\
\mathcal{L}_{\gamma
\Theta\Theta}&=&-e\bar{\Theta}^{\mu}\left[\Slash{A}+
\frac{\kappa_{\Theta}}{2M_{\Theta}}\sigma_{\nu\rho}
F^{\nu\rho}\right]\Theta_{\mu}\,+{\rm 
h.c.},\nn\\
\mathcal{L}_{\gamma
  KK^{*}}&=&g_{\gamma
  KK^{*}}\epsilon_{\mu\nu\sigma\rho}(\partial^{\mu}A^{\nu})
(\partial^{\sigma}K)K^{*\rho}\,+{\rm 
h.c.},\nn\\
\mathcal{L}_{KN\Theta}&=&\frac{g_{KN\Theta}}{M_{K}}\bar{\Theta}^{\mu}
\partial_{\mu}K{\Gamma_5}N\,+{\rm 
h.c.},\nn\\
\mathcal{L}_{K^{*}N\Theta}&=&-\frac{ig_{K^{*}N\Theta}}
{M_{V}}\bar{\Theta}^{\mu}\gamma^{\nu}[\partial_{\mu}
K^{*}_{\nu}-\partial_{\nu}K^{*}_{\mu}]\Gamma_5\gamma_5N+{\rm 
h.c.},\nn\\ \mathcal{L}_{\gamma
KN\Theta}&=&-i\frac{eg_{KN\Theta}}{M_{K}}\bar{\Theta}^{\mu}A_{\mu}K{\Gamma}_{5}N\,+{\rm
h.c.},
\label{Lagrangian}
\eee
where $N$, $\Theta^{\mu}$, $K$ and $A^{\mu}$ are the nucleon,
$\Theta^+(3/2^{\pm})$, pseudoscalar kaon and photon 
fields, respectively, and 
$\Gamma_5 = 1$ for $\Theta^+(3/2^+)$, while 
$\Gamma_5 = \gamma_5$ for $\Theta^+(3/2^-)$. In constructing the
effective Lagrangians, we assume that $\Theta^+$ has the isospin
$I=0$. Note that 
the meson-baryon couplings here are constructed in the pseudovector
(PV) scheme with the derivative acting on the kaon field.  
For the spin 3/2 case, this is the natural method to 
introduce the meson-baryon couplings including $\Theta^+$. Therefore,
we need to have the contact term as in Eq.~(\ref{Lagrangian})
explicitly.   

In Eq.~(\ref{Lagrangian}), various coupling constants are introduced 
with obvious notation. As for the electric part of the 
$\gamma\Theta\Theta$ vertex, we consider only the $g_{\mu \nu}$
term from Eq.~(5) in Ref.~\cite{Read:ye}:
$e\Theta^{\mu}\Slash{A}\Theta_{\mu}$. 
Hence,  the  
propagator of $\Theta^{\mu}(3/2^{\pm})$ is approximated to take the form of
spin-1/2 fermion.  
In fact, we have verified that the approximation 
works well since the relevant $u$-channel
contribution is strongly suppressed by the form factor. 
For the decay width of $\Theta^+$, we choose 
$\Gamma_{\Theta\to KN}=1$ MeV~\cite{Hyodo:2005wa,Eidelman:2004wy}.  
This choice gives $g_{KN\Theta}=0.53$ for
$\Theta^+(3/2^+)$ and $g_{KN\Theta}=4.22$ for
$\Theta^+(3/2^-)$. 
The unknown 
parameter $g_{K^*N\Theta}$ is estimated by the quark model. 
As for $\Theta^+(3/2^+)$ we employ the relation
$|g_{K^*N\Theta}|=\sqrt{3}g_{KN\Theta}$~\cite{Close:2004tp}, which 
is applicable to both $1/2^+$ and $3/2^+$ states.  
On the other hand, we find $|g_{K^*N\Theta}|\sim 2$ for
$\Theta^+(3/2^-)$ of $(0s)^5$ configuration. 
In numerical calculation, we test the values of $\pm
|g_{K^*N\Theta}|$ and $0$ in order to see the role of $K^*$. 

As for the form factor, we employ the four dimensional gauge and
Lorentz invariant one which was used  in Ref.~\cite{Nam:2005uq}. 
There, it was shown that this form factor with 
the cutoff $\Lambda=750$ MeV 
reproduced experimental data of the 
photoproduction of $\Lambda(1520)$~\cite{Nam:2005uq} qualitatively well. 
As for the anomalous magnetic moment of $\Theta^+$, we choose 
$\kappa_{\Theta} = 1$, which is an upper bound among the 
existing estimation.  
In the quark model, its value turns out to be very small
if $\Theta^+$ has $J^P=3/2^-$~\cite{Nam:2005uq}. 
The resulting amplitude, however, does not depend much on $\kappa_{\Theta}$, 
since the form factor suppresses the $s$- and $u$-channel contributions
that are proportional to $\kappa_{\Theta}$.  
Since the calculation of the Born terms for $\Theta^+(3/2^{\pm})$
photoproduction is analogous to that of $\Lambda^*$, we refer
to Ref.~\cite{Nam:2005uq} for details.     

\begin{figure}[tbh]
\begin{tabular}{cc}
\resizebox{4.5cm}{4.5cm}{\includegraphics{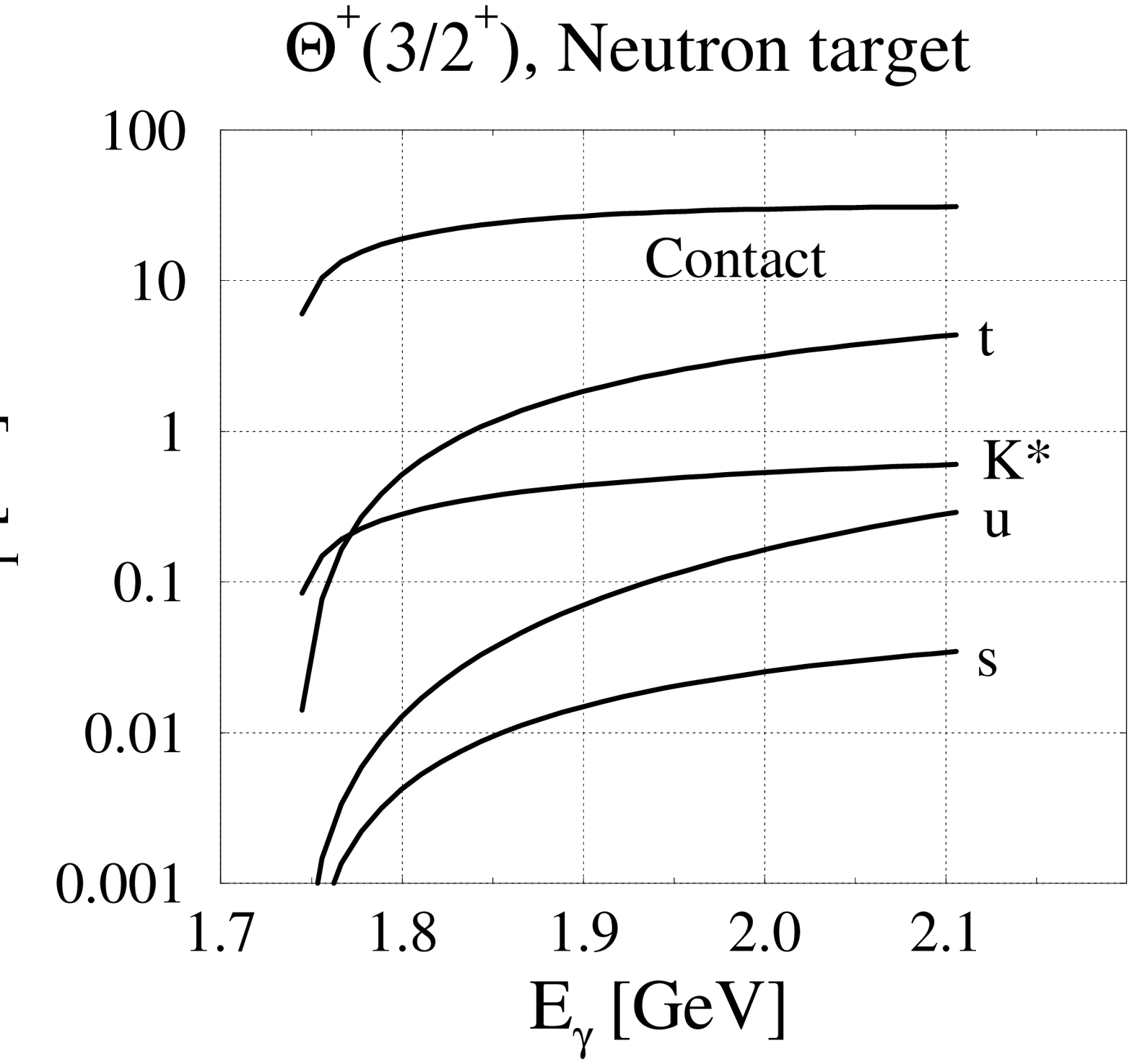}}
\resizebox{4.1cm}{4.5cm}{\includegraphics{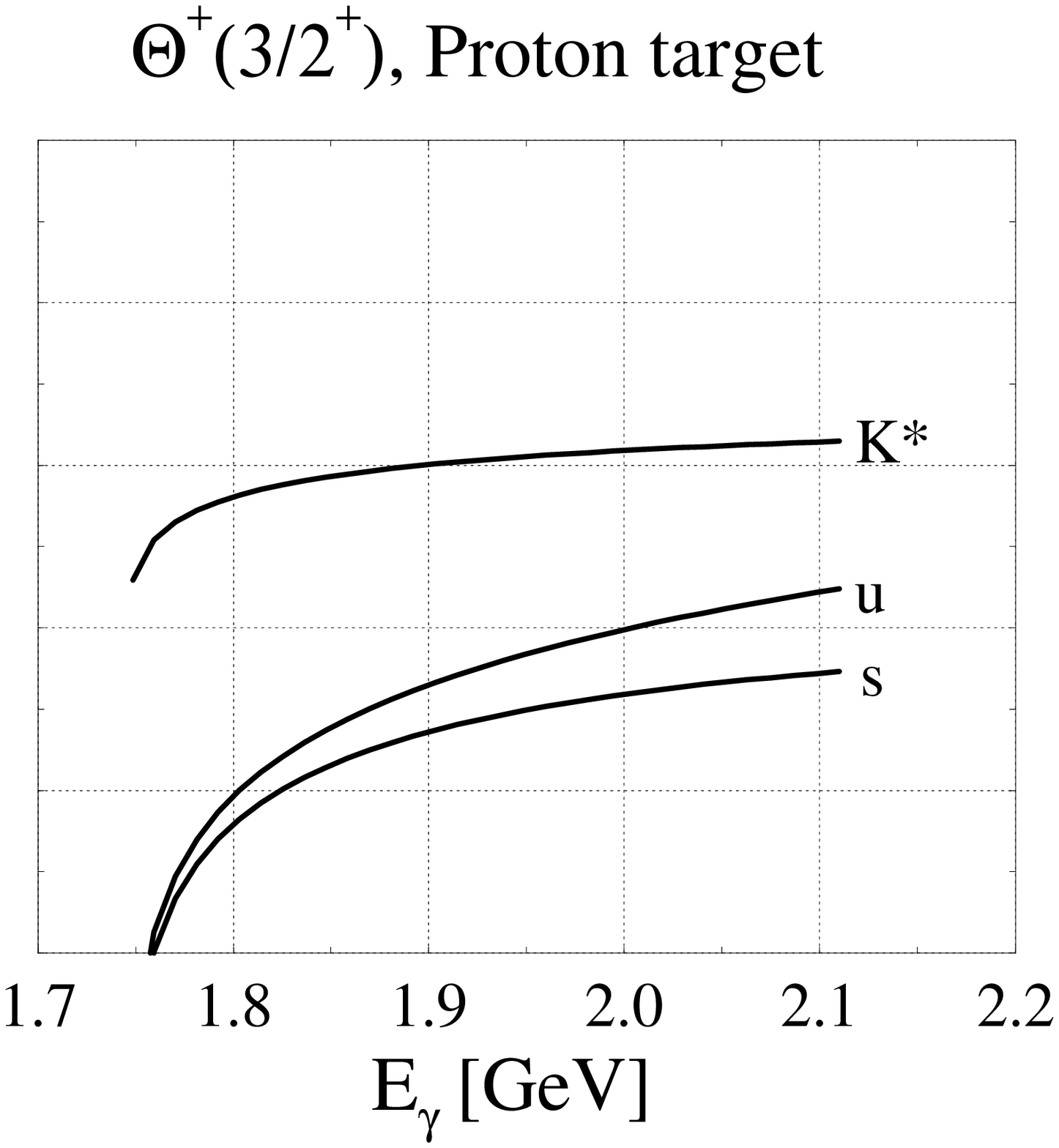}}
\end{tabular}
\begin{tabular}{cc}
\end{tabular}
\begin{tabular}{cc}
\resizebox{4.5cm}{4.5cm}{\includegraphics{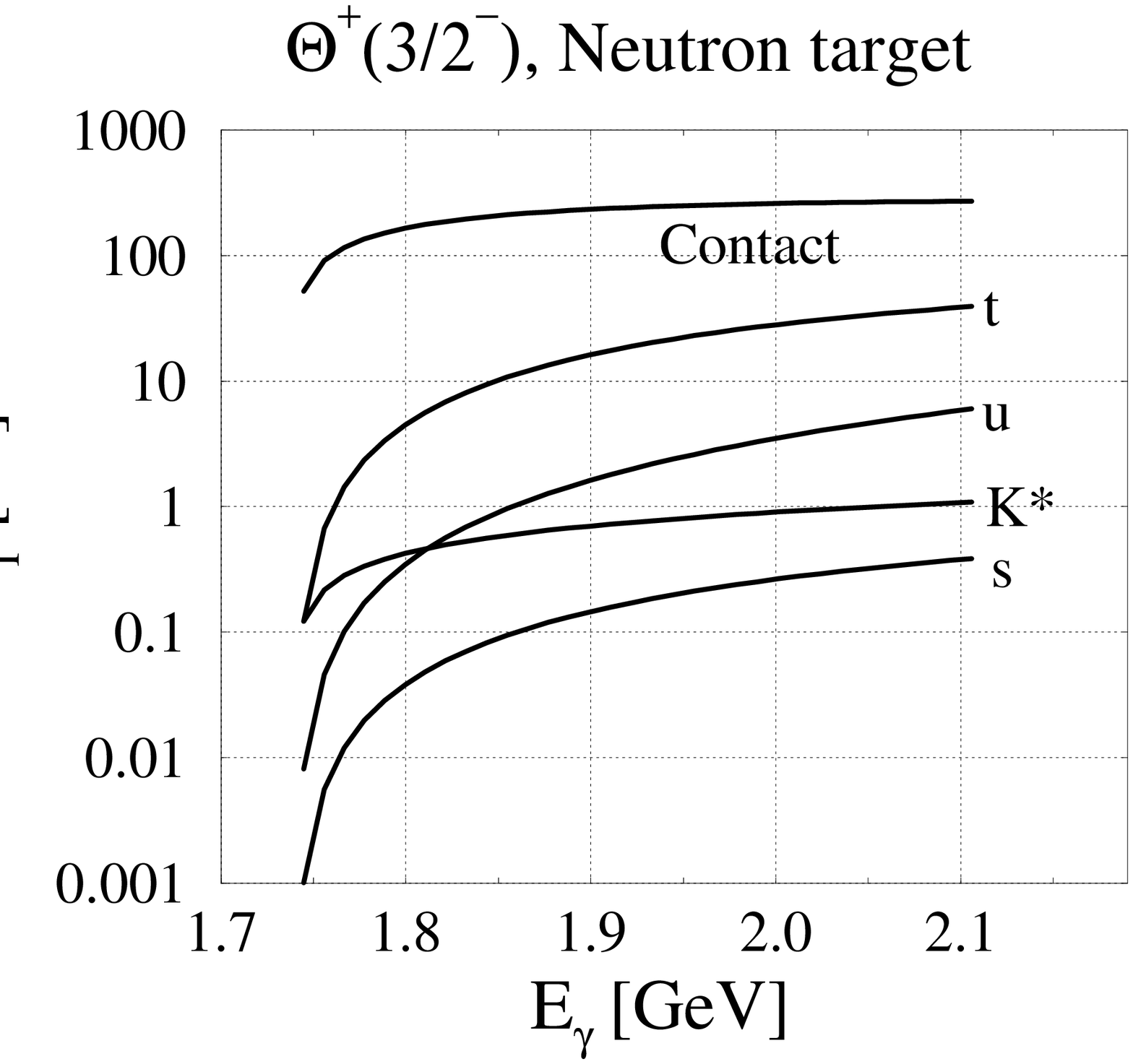}}
\resizebox{4.1cm}{4.5cm}{\includegraphics{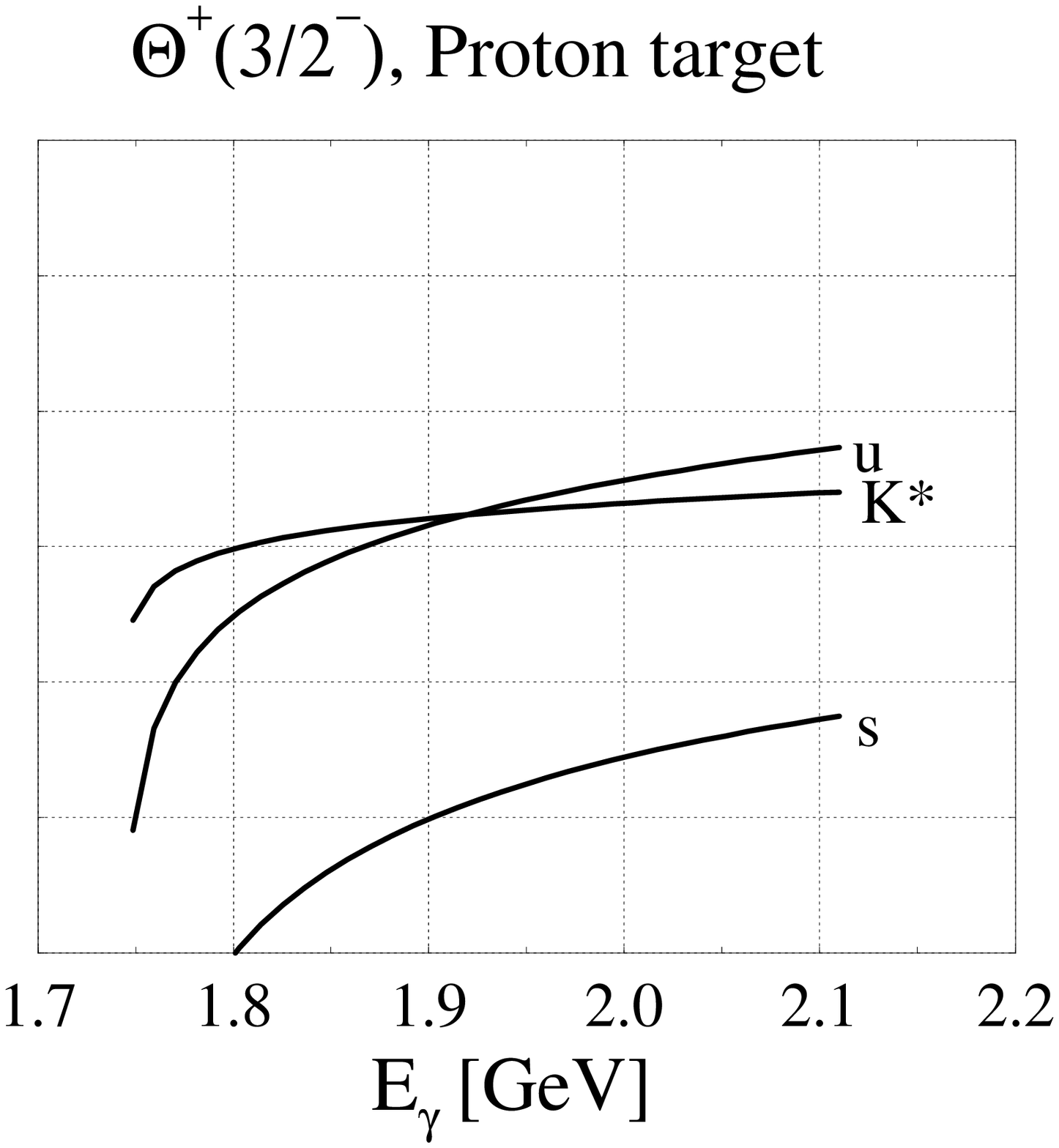}}
\end{tabular}
\caption{Total cross sections for each kinematical channel for
$J^P=3/2^+$ (upper two panels) and for $J^P=3/2^-$ (lower two
panels). Here, we use $\Gamma_{\Theta\to KN}=1.0$
MeV. $g_{K^*N\Theta}=+0.91$ for the positive parity and $+2$ for the
negative one.}    
\label{fig0}
\end{figure}

Let us now discuss our results.  
First, we show various contributions to the total cross section 
in Fig.~\ref{fig0}, where results are 
shown separately as functions of the incident photon energy $E_\gamma$
in the laboratory frame for  
the $s$-, $t$-, $u$-channels, $K^*$-exchange and the contact term for
the neutron target, and for $s$-, $u$-channels, and $K^*$-exchange for the
proton.  It is shown that the largest contribution comes from the
contact term which is present only for the neutron, while the $u$- and
$s$-channel contributions are strongly suppressed due to the form
factor.  This is so because the baryon in the $u$- and $s$-channels
are  further off mass shell than in the $t$-channel.   
The $K^*$-exchange contributes some, but the amount 
is significantly smaller than that of the contact term. From these
observations, we expect that  
whether the contact term is present or not 
yields the large difference in the production rates from the proton 
and neutron targets.  

\begin{figure}[tbh]
\begin{tabular}{cc}
\resizebox{4.5cm}{4.5cm}{\includegraphics{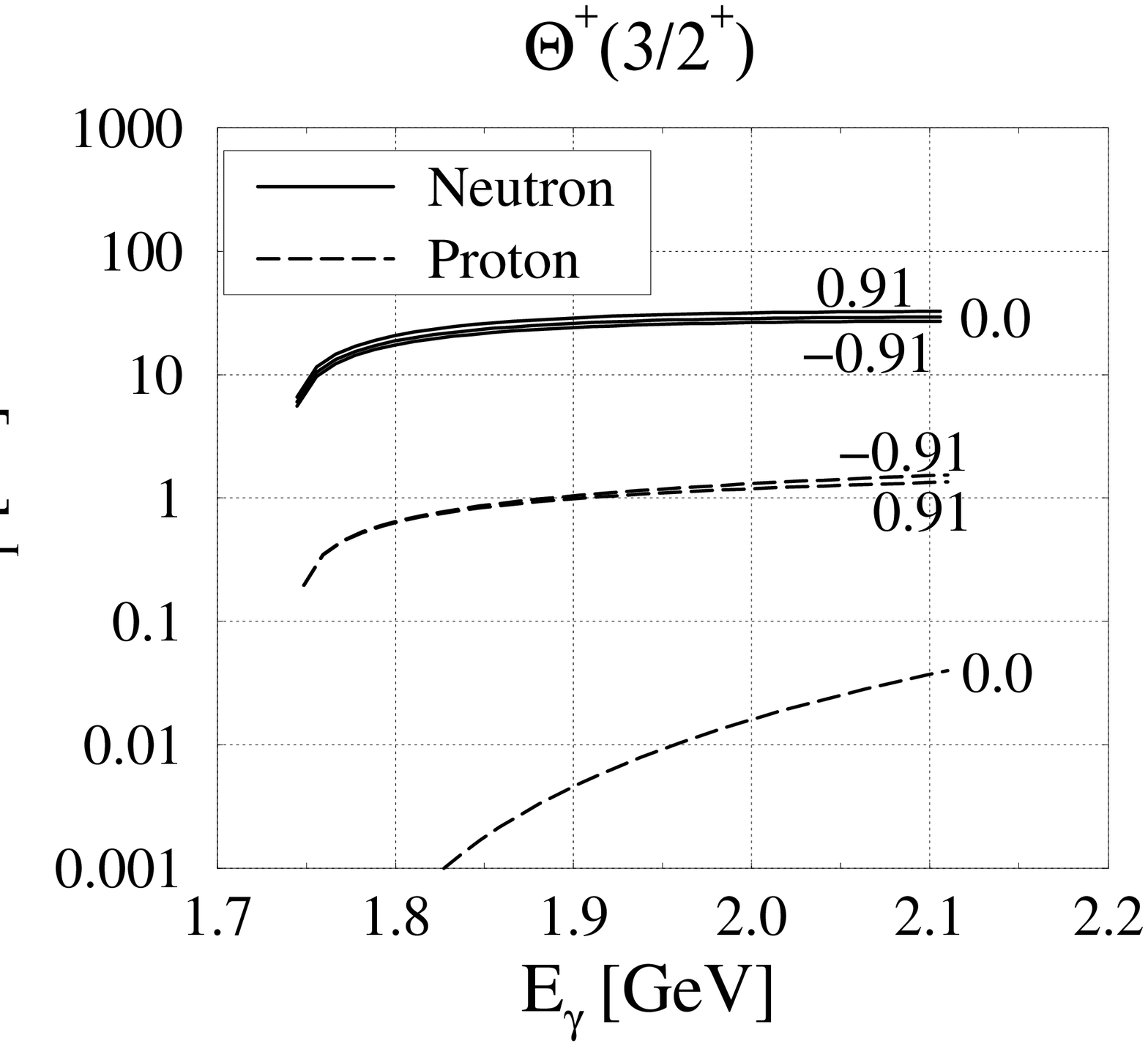}}
\resizebox{4.1cm}{4.5cm}{\includegraphics{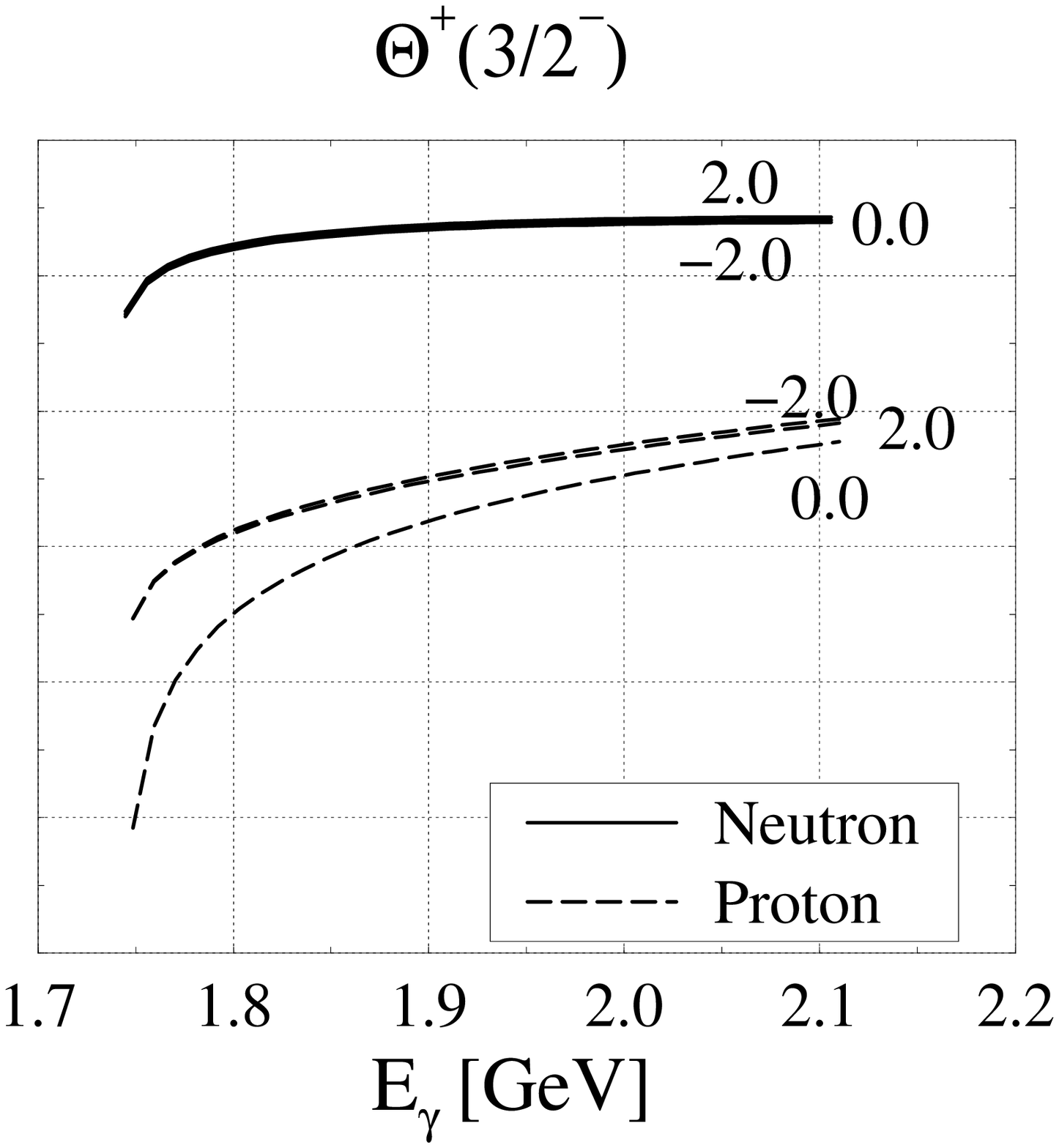}}
\end{tabular}
\begin{tabular}{cc}
\end{tabular}
\begin{tabular}{cc}
\resizebox{4.5cm}{4.5cm}{\includegraphics{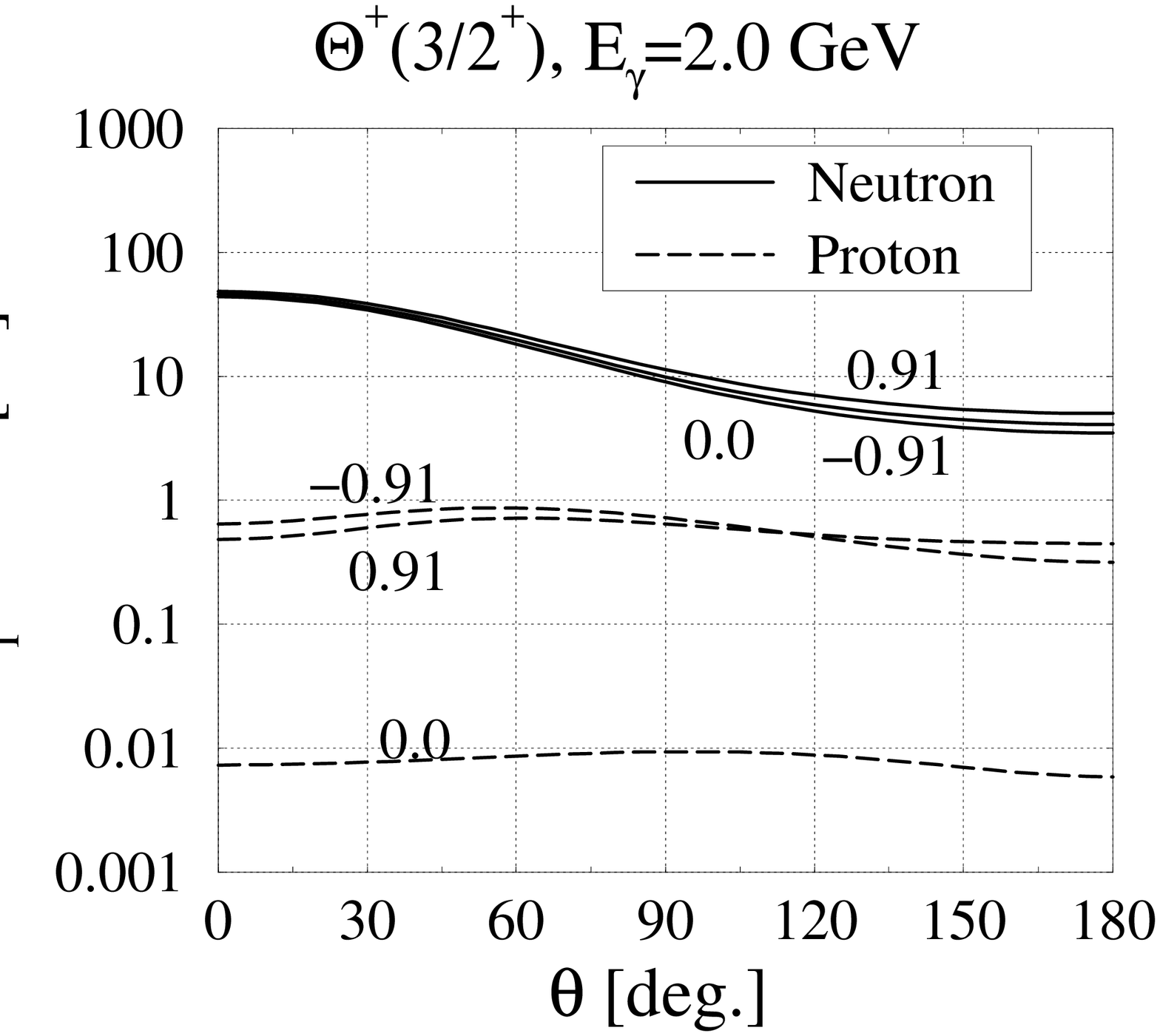}}
\resizebox{4.1cm}{4.5cm}{\includegraphics{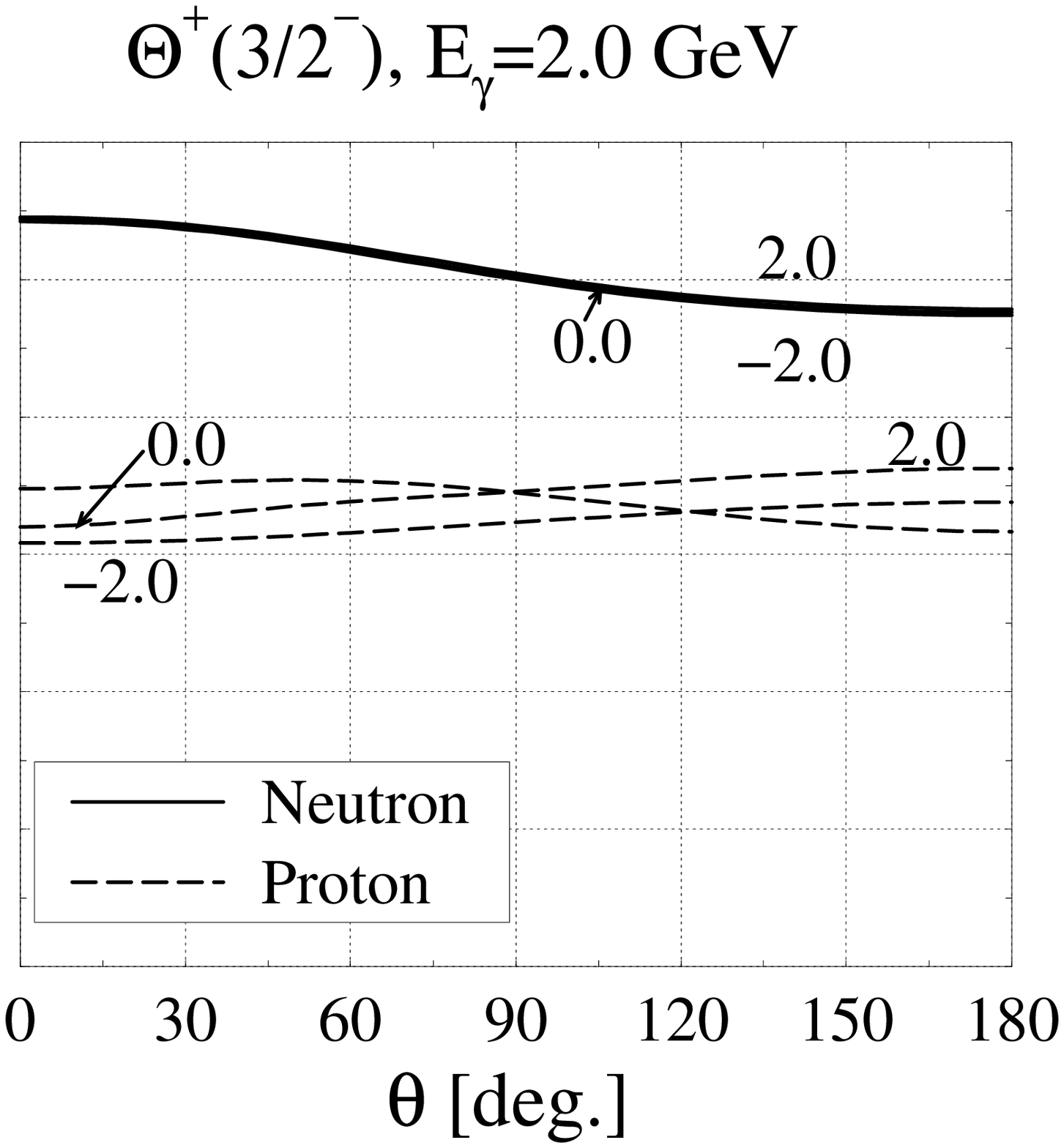}}
\end{tabular}
\caption{Upper two panels: Total cross sections for $J^P=3/2^+$ (left) and
for $J^P=3/2^-$ (right). Lower two panels: Differential cross
sections for $J^P=3/2^+$ (left) and for  
$J^P=3/2^-$ (right). The numbers on the figures denote the values of
the coupling constant $g_{K^*N\Theta}$.}
\label{fig1}
\end{figure}

We can verify this explicitly 
as shown in the upper two panels of 
Fig.~\ref{fig1}, where 
the total cross sections including all kinematical channels 
are plotted in logarithmic scale 
for $J^P = 3/2^+$ (left panel) and $3/2^-$ (right panel),  
and for proton (dashed curves) and neutron (solid curves).  
Three curves are obtained by using different 
values of $g_{K^*N\Theta}$ as indicated in the figures.  
As anticipated, the contact term is dominant for the neutron 
target, which makes the cross sections significantly 
larger than for the proton target.  
The cross sections averaged in the energy range 
$1.73 < E_{\gamma} < 2.10$ GeV are summarized in Table~\ref{table1}.  
The numerical values are evaluated when 
$|g_{K^*N\Theta}| = 0.91 (J^P = 3/2^+)$ 
and 2 $(J^P = 3/2^-)$: They do not depend much on the sign 
of the coupling constant as we can see from Fig.~\ref{fig1}. 
We find that the $\Theta^+$ production rate 
is larger for the neutron than for the proton, 
when the above finite values are used for $g_{K^*N\Theta}$, 
by factors $\sim25\,(3/2^+)$ and $\sim50\,(3/2^-)$. 
If $g_{K^*N\Theta} = 0$, the difference is even more 
enhanced.  

Absolute values are larger for the negative 
parity than for the positive one, 
due to the large difference in the coupling constant;
$g_{KN\Theta}(J^P = 3/2^+) = 0.53$ and 
$g_{KN\Theta}(J^P = 3/2^-) = 4.22$.  
This difference stems from the different coupling structure
of $\Theta^+$ decaying into the $KN$ channel, i.e.
$p$-wave for $3/2^+$ and $d$-wave for $3/2^-$~\cite{He:2004tj}.  

Once again we note that 
these cross sections are computed by using the parameters 
corresponding to $\Gamma_{\Theta\to KN}=1.0$.  
For other values of $\Gamma_{\Theta\to KN}$, 
they are precisely proportional to $\Gamma_{\Theta\to KN}$
for $3/2^+$, while it is approximate for $3/2^-$.

In the lower two panels of 
Fig.~\ref{fig1} we present the differential cross
sections when $E_{\gamma}=2.0$ GeV. There, $\theta$ is defined by the
angle between the incident photon and the outgoing kaon in the center
of mass frame. For the neutron target 
we observe strong forward enhancement, which is 
the characteristic feature of the contact term with 
the gauge invariant form factor.   
On the contrary, for the proton target a
bump appears at around $60^{\circ}$ when $K^*$-exchange 
is present; the angle dependence of the 
$K^*N\Theta$ coupling characterizes it.  
If the $K^*$-exchange is absent, 
the cross section shows a backward peak which is the typical behavior
of the $u$-channel process.  These different angular distributions may
help understand the production mechanism of $\Theta^+$.  

Now it is interesting to compare the above results of 
$J^P = 3/2^{\pm}$ with those of $J^P = 1/2^+$.  
Here we assume the positive parity, since the negative parity 
is not likely to be compatible with the narrow decay 
width~\cite{Hosaka:2004bn}.  
The reaction for the $\Theta^+$ of $J^P = 1/2^+$ has been 
investigated previously and we refer to 
Refs.~\cite{Nam:2003uf,Nam:2004xt} 
for details. 
An important fact for the case of $J^P = 1/2^+$ is that one can 
construct the effective Lagrangians both in the pseudoscalar (PS) and 
in the pseudovector (PV) schemes.  
In a consistent description, these two schemes should be
equivalent in the strong interaction sector.  
In fact, it was shown explicitly that the difference between the 
two schemes are due to the terms of anomalous magnetic 
moments which is the electromagnetic coupling~\cite{Nam:2003uf,Nam:2004xt}.  
Hence for the computation of cross sections 
we shall work out in the PS scheme with using the 
four dimensional form factor.  
To make our comparison fair, we adopt the parameters 
reproducing  
$\Gamma_{\Theta\to KN}=1.0$ MeV~\cite{Eidelman:2004wy} 
which gives $g_{KN\Theta}=1.0$ and
$|g_{K^*N\Theta}| = \sqrt{3}g_{KN\Theta}=1.73$.  
For the magnetic moment we have used $\kappa_{\Theta} = 1$ 
again.  

\begin{figure}[tbh]
\begin{tabular}{cc}
\resizebox{4.5cm}{4.5cm}{\includegraphics{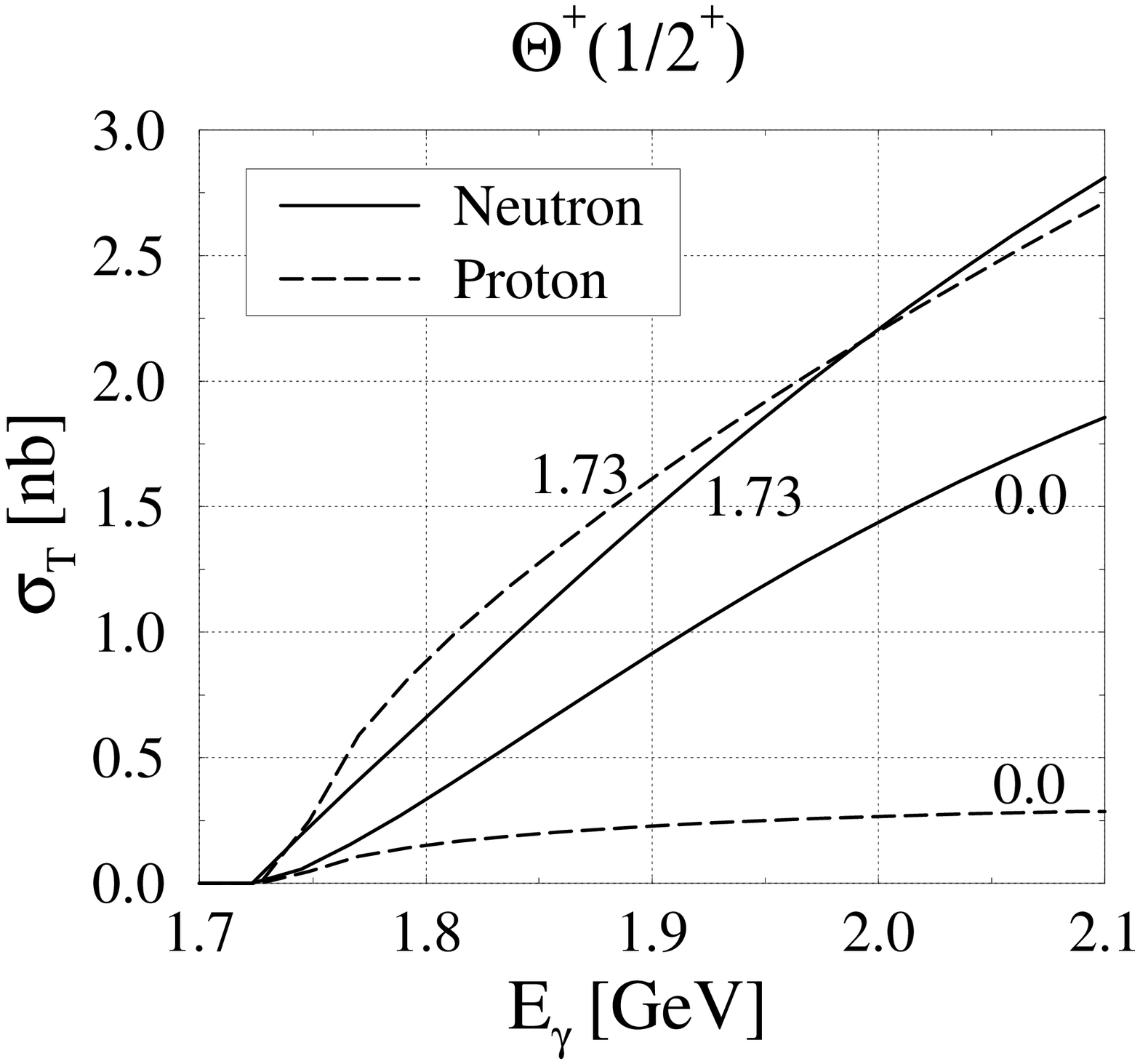}}
\resizebox{4.1cm}{4.5cm}{\includegraphics{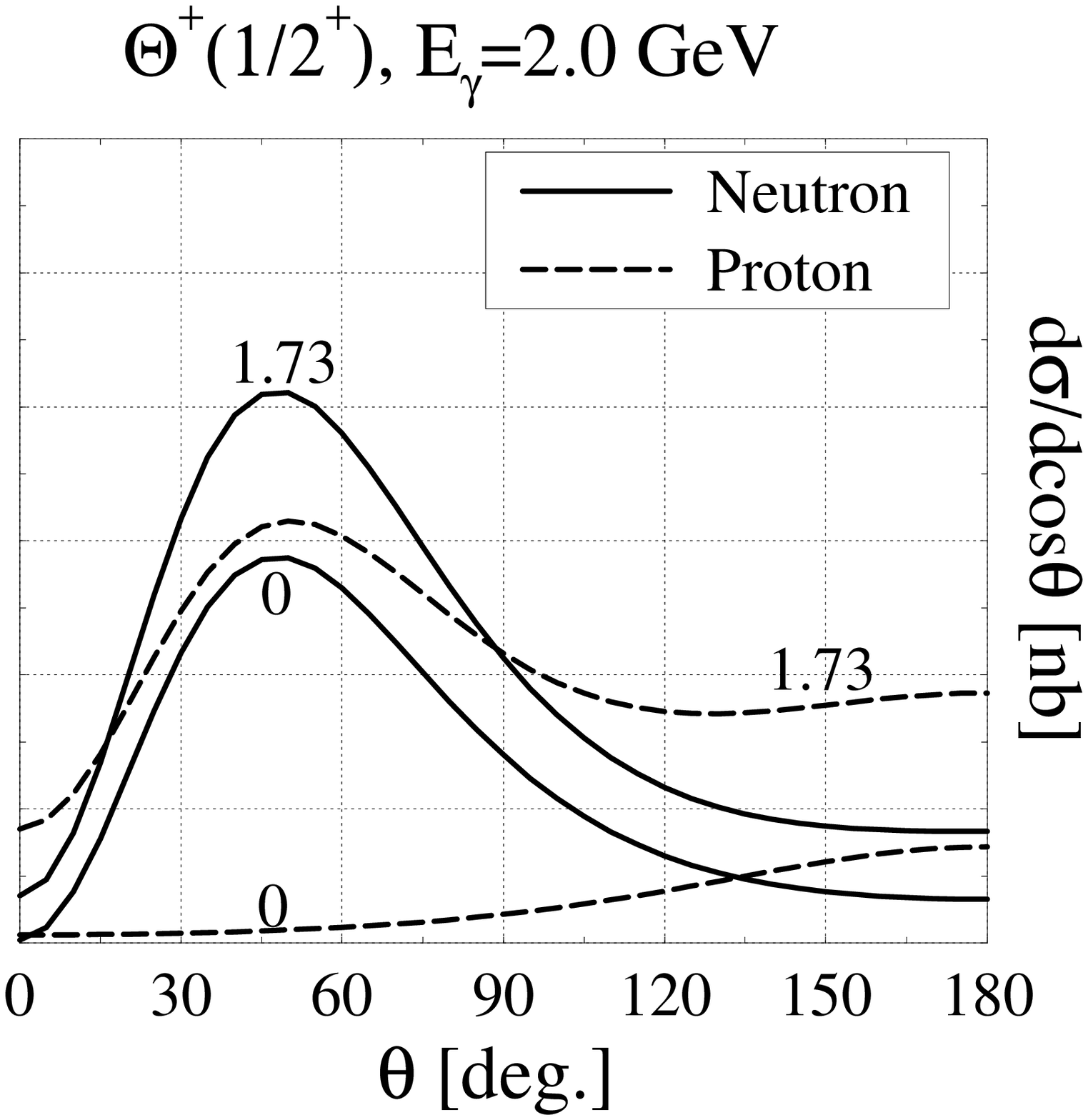}}
\end{tabular}
\caption{Total (left) and differential (right) cross sections for
$\Theta^+(1/2^+)$. The numbers on the figures denote the values of the
coupling constant $g_{K^*N\Theta}$.}
\label{fig2}
\end{figure}

In the left panel of Fig.~\ref{fig2}, we show the total cross
sections as functions of $E_\gamma$, where we show the results with 
$g_{K^*N\Theta} = 0$ and +1.73.  
The result with $g_{K^*N\Theta} = -1.73$ is qualitatively 
similar to that of $g_{K^*N\Theta} = +1.73$.  
When $g_{K^*N\Theta}$ has a sizable value 
($g_{K^*N\Theta}= +1.73$), 
the total cross sections for the two different targets do not
show obvious differences.  
On the other hand, the production rate is 
suppressed by a factor $\lsim$ 5 for the proton
target, when $K^*$-exchange is absent.  
In both cases the cross sections of the proton is less suppressed 
as compared to the neutron, a feature which is rather
different from the case of $\Theta^+(J^P = 3/2^{\pm})$. 
We can understand this from the different role of the contact term.  
To this end, we consider the results in the PV scheme, where 
the contact term appears necessarily.  
{}For $J^P = 1/2^+$, the gauge invariance and the equivalence 
between the PS and PV schemes requires that the form factor 
of the contact term must be the same as for the $u$-channel.  
Consequently, the contributions of  
the $u$-channel and contact term are similar.  
In contrast, for the spin 3/2 cases where
only the PV scheme is available, the contact term 
remains dominant.  

The differential cross sections at $E_{\gamma}=2.0$ GeV are shown in
the right panel of Fig.~\ref{fig2}. 
In three cases curves show a bump at around 
$45^{\circ}$, which is a feature of both the 
$K$-exchange and $K^*$-exchange terms.  
Only when both of them are absent (for the proton without $K^*$), 
the bump structure disappears, where instead, the backward peak
appears as is the property of the $u$-channel contribution.  
These angular distributions are once again very much different from 
the case of $J^P = 3/2^{\pm}$ in particular for 
the neutron target.  

In the present work, we have studied the photoproduction of
$\Theta^+$ for $J^P=3/2^{\pm}$ and $1/2^+$, where 
we have computed the Born diagrams using the effective
Lagrangians.  
The Rarita-Schwinger formalism was utilized for the description of  
spin 3/2 $\Theta^+$. 
The gauge and Lorentz invariant four dimensional   
form factor was employed with the cutoff determined to 
reproduce the data of the 
$\Lambda(1520)$ photoproduction~\cite{Nam:2005uq}.  
Several unknown parameters were
estimated by the quark model and some phenomenological considerations.  

We have then found that the production rate for the proton target is 
significantly suppressed as compared to that of the neutron.  
This result is governed by the 
contact term which is present only in the charge exchange 
process for the neutron target.  
A similar observation was made in the previous study of the 
photoproduction of $\Lambda(1520)$, where the role of the 
proton and neutron was interchanged.  
The present result was obtained 
by fixing several parameters in an expectedly reasonable manner, 
especially by using the form factors determined 
in the phenomenological 
analysis of the $\Lambda(1520)$ photoproduction, 
hoping that the form factors for $\Theta^+$ and 
$\Lambda(1520)$ are not very much different. It is therefore important
to check experimentally the asymmetric nature in the photoproduction
of $\Lambda(1520)$. An experiment with the deuteron target was
recently performed by the LEPS collaboration, and the analysis is
currently performed~\cite{nakano}. 

In Table~\ref{table1}, we summarize and compare various 
cross sections averaged in the energy range 
$1.73 < E_{\gamma} < 2.10$ GeV.  From the table, in all cases, the
cross sections  
for the proton target are of order of a few nb.  
Interestingly, these values seem compatible with the 
upper bound, 
if $\Theta^+$ exists, as
extracted from the recent CLAS experiment~\cite{DeVita:2005CLAS}.  
Therefore, one of our conclusions is that the 
results of the CLAS does not immediately lead to 
the absence of $\Theta^+$.  
Our present study have shown 
the photoproduction of 
$\Theta^+$ could be suppressed for the proton target.  
In contrast, the cross sections are sizable 
for the neutron target as large as 
25 nb ($J^P = 3/2^+$) and 
200 nb ($J^P = 3/2^-$), 
when $\Gamma_{\Theta \to K N} = 1$ MeV is employed.   
The future experimental data from the neutron target 
with higher statistics with information on 
cross section values as well as angular distributions 
are extremely important to settle the issue of the pentaquark 
$\Theta^+$.\\
\\  

{\bf Acknowledgment}\\
We are very grateful to J.~K.~Ahn, K.~Hicks, T.~Hyodo, T.~Nakano, A.~Titov and
H.~Toki for fruitful discussions.  The work of S.I.N. has been supported by 
the scholarship from the Ministry of Education, Culture,
Science and Technology of Japan. The work of A.H. is also supported in
part by the Grant for Scientific Research ((C) No.16540252) from the
Education, Culture, Science and Technology of Japan. The works of
H.C.K. and S.I.N. are supported by the Korean Research Foundation
(KRF--2003--070--C00015).     

\begin{table}[tbh]
\begin{tabular}{|c|c|c|c|c|c|c|}
\hline
$J^P$&\multicolumn{2}{c|}{$3/2^+$}&
\multicolumn{2}{c|}{$3/2^-$}&\multicolumn{2}{c|}{$1/2^+$}\\
\hline
$g_{KN\Theta}$&\multicolumn{2}{c|}{$0.53$}&
\multicolumn{2}{c|}{$4.22$}&\multicolumn{2}{c|}{$1.0$}\\ 
\hline
$g_{K^*N\Theta}$&\multicolumn{2}{c|}{$\pm0.91$}&
\multicolumn{2}{c|}{$\pm2$}&\multicolumn{2}{c|}{$\pm1.73$}\\ 
\hline
Target&$n$&$p$&$n$&$p$&$n$&$p$\\
\hline
$\sigma$&$\sim25$ nb&$\sim 1$ nb&$\sim 200$ nb&$\sim4$ nb&$\sim1$
nb&$\sim1$ nb\\    
$\frac{d\sigma}{d\cos\theta}$&Forward&$\sim60^{\circ}$&Forward&--&$\sim45^{\circ}$&$\sim45^{\circ}$\\ 
\hline
\end{tabular}
\caption{Main results of the $\Theta^+$ photoproduction, where all 
results are for the case of finite $g_{K^*N\Theta}$. }
\label{table1}
\end{table}

 
\end{document}